# Heavy-ion double charge exchange reactions: a tool towards *0υββ* nuclear matrix elements

F. Cappuzzello[a,b], M. Cavallaro[b*], C. Agodi[b], M. Bondì[a,b], D. Carbone[b], A. Cunsolo[b], A. Foti[a,c]

[a] Dipartimento di Fisica e Astronomia, Università di Catania, Catania, Italy
[b] INFN - Laboratori Nazionali del Sud, Catania, Italy
[c] INFN - Sezione di Catania, Catania, Italy

*Corresponding author: manuela.cavallaro@lns.infn.it

**Abstract.** The knowledge of the nuclear matrix elements for the neutrinoless double beta decay is fundamental for neutrino physics. In this paper, an innovative technique to extract information on the nuclear matrix elements by measuring the cross section of a double charge exchange nuclear reaction is proposed. The basic point is that the initial and final state wave functions in the two processes are the same and the transition operators are similar. The double charge exchange cross sections can be factorized in a nuclear structure term containing the matrix elements and a nuclear reaction factor. First pioneering experimental results for the $^{40}$Ca($^{18}$O,$^{18}$Ne)$^{40}$Ar reaction at 270 MeV incident energy show that such cross section factorization reasonably holds for the crucial $0^+ \to 0^+$ transition to $^{40}$Ar$_{gs}$, at least at very forward angles.

## 1. Introduction

Neutrinoless double beta decay (*0νββ*) is potentially the best resource to probe the Majorana or Dirac nature of neutrino and to extract its effective mass. Moreover, if observed, *0νββ* will signal that the total lepton number is not conserved. Presently, this physics case is one of the most important researches "beyond the Standard Model" and might guide the way toward a Grand Unified Theory of fundamental interactions.

Since the *ββ* decay process involves transitions in atomic nuclei, nuclear structure issues must be accounted for to describe it. The 0*νββ* half-life $T_{1/2}$ can be factorized as a phase-space factor $G_{0\nu}$, the nuclear matrix element (NME) $M_{0\nu}$ and a term $f(m_i, U_{ei})$ containing the masses $m_i$ and the mixing coefficients $U_{ei}$ of the neutrino species:

$$[T_{1/2}]^{-1} = G_{0\nu} |M_{0\nu}|^2 |f(m_i, U_{ei})|^2 \qquad (1)$$

where the NME is the transition amplitude from the initial $\varphi_i$ to the final $\varphi_f$ nuclear state of the *ββ* process through the 0ν*ββ* decay operator:

$$|M_{0\nu}|^2 = \left| \left\langle \varphi_f \left| O^{0\nu\beta\beta} \right| \varphi_i \right\rangle \right|^2 \qquad (2)$$

Thus, if the NMEs are established with sufficient precision, the neutrino masses and the mixing coefficients can be extracted from 0*νββ* decay rate measurements.

The evaluation of the NMEs is presently limited to state of the art model calculations based on different methods (QRPA, shell-model, IBM, EDF, etc.) [1], [2], [3], [4]. High precision experimental information from single charge exchange (CE), transfer reactions and electron capture are used to constraint the calculations [5], [6], [7], [8], [9]. However, the ambiguities in the models are still too large and the constraints too loose to reach accurate values of the NMEs. Discrepancy factors higher than two are presently reported in literature [10]. In addition some assumptions, common to the different competing calculations, could cause overall systematic uncertainties [11].

The experimental study of other nuclear transitions where the nuclear charge is changed by two units leaving the mass number unvaried, in analogy to the $\beta\beta$-decay, could give important information. Past attempts to use pion double charge exchange reactions [12], [13], [14], [15] to probe $\beta\beta$-decay NMEs were abandoned due to the large differences in the momentum transfers and in the nature of the operators [11]. Early studies of heavy-ion induced double charge exchange reactions (DCE) were also not conclusive. The reason was the lack of zero-degree data and the poor yields in the measured energy spectra and angular distributions, due to the very low cross sections involved, ranging from about 5-40 nb/sr [16], [17] to 10 µb/sr [18]. Actually, this wide range of observed cross sections has never been deeply discussed. An additional complication in the interpretation of the data was due to possible contributions of multi-nucleon transfer reactions leading to the same final states [19], [20], [21].

Here we show that the use of modern high resolution and large acceptance spectrometers allows to face the experimental challenges and to extract quantitative information from DCE reactions. The measurement of DCE high-resolution energy spectra and accurate cross sections at very forward angles is crucial to identify the transitions of interest [22]. The concurrent measurement of the other relevant reaction channels allows isolating the direct DCE mechanism from the competing transfer processes. These are at least of $4^{th}$-order and can be effectively minimized by the choice of the proper projectile-target system and incident energy [23].

## 2. DCE reactions and $0\nu\beta\beta$ decays

The availability for the first time of valuable data on DCE reactions raises the question whether they can be used toward the experimental access to $0\nu\beta\beta$ NMEs. Although the DCE and $0\nu\beta\beta$ decay processes are mediated by different interactions, there are a number of important similarities among them:

    1. Parent/daughter states of the $0\nu\beta\beta$ decay are the same as those of the target/residual nuclei in the DCE;

    2. Short-range Fermi, Gamow-Teller and rank-2 tensor components are present in both the transition operators, with relative weight depending on incident energy in DCE;

    3. A large linear momentum (~100 MeV/c) is available in the virtual intermediate channel in both processes [10]. This is a crucial similarity since other processes cannot probe this feature [24] while muon capture can only from one side of $0\nu\beta\beta$ transition [25], [26].

    4. The two processes are non-local and are characterized by two vertices localized in a pair of valence nucleons;

5. Both processes take place in the same nuclear medium. In medium effects are expected to be present in both cases, so DCE data could give a valuable constraint on the theoretical determination of quenching phenomena on *0νββ*. One should mention for example that in single *β*-decay, *2νββ*-decay [4] and charge exchange reactions [27], the limited model space used in the calculations and the contribution of non-nucleonic degrees of freedom and other correlations require a renormalization of the coupling constants in the spin-isospin channel. However an accurate description of quenching has not yet been fully established and other aspects of the problem can give important contributions [28];

6. An off-shell propagation through virtual intermediate channels is present in the two cases. The virtual states do not represent the asymptotic channels of the reaction and their energies can be different from those (measurable) at stationary conditions [29]. In practice, a supplementary contribution of several MeV to the line width is present in the intermediate virtual states. This is related to the transit time of a particle (neutrino in one case and pair of nucleons in the other) along the distance of the two vertices of the *0νββ* and DCE processes. The situation is very different in CE reactions, where the intermediate states of *0νββ* are populated as stationary ones and in *2νββ*, where the neutrinos and electrons are projected out from the nucleus. No effective broadening of the line width is thus probed in CE and *2νββ*.

The description of NMEs for DCE and *0νββ* presents the same degree of complexity, with the advantage for DCE to be "accessible" in laboratory. However, a simple relation between DCE cross sections and *ββ*-decay half-lives is not trivial and needs to be explored. In particular the DCE reaction is, to its leading order, a two-step process involving projectile and target internal structure as well as the full nucleus-nucleus interaction. A coherent approach to this problem is in principle the multistep direct nuclear reaction theory, extensively developed in different quantum-mechanical frameworks [30], [31], [32], [33] and implemented in specific codes [34], [35]. However, none of these is, to our knowledge, developed for DCE reactions. For this reason in the following we will concentrate on a simpler two-step approach which contains all the relevant features and provides at least an approximate estimate of the nuclear matrix elements.

## 3. Factorization of DCE cross section

It is well known that single *β*-decay strengths are proportional to CE reaction cross sections for linear momentum transfer $q \sim 0$ and under specific conditions [27], [36], [37], [38], [39]:

$$\frac{d\sigma}{d\Omega}(q,E_x) = \hat{\sigma}_\alpha(E_p,A) F_\alpha(q,E_x) B_T(\alpha) B_P(\alpha) \qquad (3)$$

where $B_T(\alpha)$ and $B_P(\alpha)$ are the target and projectile *β*-decay reduced transition strengths (related to the matrix elements $M(\alpha)$[1]) for the $\alpha$ = Fermi (*F*) or Gamow-Teller (*GT*) operators[2]. The factor $F_\alpha(q,E_x)$ describes the shape of the cross section distribution as a function of the linear momentum transfer $q$ and the excitation energy $E_x$. For $L = 0$ transitions, it depends on the square of the $j_0(qr)$ spherical Bessel function [37], [27]. The quantity $\hat{\sigma}_\alpha$, named "unit cross section", is of primary interest since it almost behaves as a universal property of the nuclear response to $F$ and $GT$ probes. The dependence on the projectile energy $E_p$ and on the target mass number $A$ is in fact quite smooth and computable all along the nuclear chart. In a rigorous Distorted Wave approach as that proposed by Taddeucci et al. [37], the unit cross section for a CE process is factorized as:

$$\hat{\sigma}(E_p,A) = K(E_p,0)|J_\alpha|^2 N_\alpha \qquad (4)$$

where $K(E_p,E_x)$ is a kinematic factor, $J_\alpha$ is the volume integral of the effective isovector nucleon-nucleon interaction and $N_\alpha$ expresses the distortion of the incoming and outcoming waves in the scattering [39].

---

[1] In this paper $B(\alpha) = \frac{1}{2J_i+1}|M(\alpha)|^2$, where $J_i$ is the angular momentum of the initial state.
[2] Usually for (p,n) and (n,p) reactions the $B_P(\alpha)$ strength does not explicitly appear in the formula and is included in $\hat{\sigma}_\alpha(E_p,A)$. In this paper, following this convention, $B_P(\alpha)$ in eq. (3) is divided by the $B_P(\alpha)$ related to the (p,n), which is 3.049 [40].

Eqs. (3)-(4) are routinely used for accurate (within few percent) determination of the strengths $B$ in light-ion induced reactions such as (n,p), (p,n), ($^3$He,t), (t,$^3$He), (d,$^2$He) at bombarding energies above 100 AMeV [40], [24], [41], [42]. For heavy-ion induced reactions, the data analyses are typically more involved, due to the projectile degrees of freedom and the sizeable amount of momentum transfer. In addition, the contribution of sequential nucleon exchange to CE cross section should also be considered [43], [44], [45], [46]. A relevant simplification comes from the strong absorption of the scattering waves in the inner part of the colliding systems and the resulting surface localization of such reactions. As a consequence in these cases, the use of fully consistent microscopic approaches with double folded potentials for the reaction form factors still allows the determination of $B(\alpha)$ within 10-20% [45].

Under the hypothesis of a surface localized process, a generalized version of eq. (3) is assumed also for DCE within a similar distorted wave approach:

$$\frac{d\sigma^{DCE}}{d\Omega}(q,E_x) = \hat{\sigma}_\alpha^{DCE}(E_p,A) F_\alpha^{DCE}(q,E_x) B_T^{DCE}(\alpha) B_P^{DCE}(\alpha) \qquad (5)$$

where the superscripts indicate that the factors refer to the DCE process. The $B_{T,P}^{DCE}(\alpha)$ are connected to the nuclear matrix elements of the $\beta\beta$-decay. In analogy to the CE, the $q$ dependence of the cross section is given by a Bessel function. A DCE unit cross section can be defined as follows:

$$\hat{\sigma}_\alpha^{DCE}(E_p,A) = K^{DCE}(E_p,0)|J_\alpha^{DCE}|^2 N_\alpha^{DCE} \qquad (6)$$

where $K^{DCE}(E_p,0)$ and $N_\alpha^{DCE}$ are analogous to eq. (4) and $J_\alpha^{DCE}$ is the volume integral of the double charge exchange interaction. A closer inspection of eqs. (4) and (6) reveals that the specificity of the single or double charge exchange unit cross sections is expressed through the volume integrals of the potentials, while the other factors are general features of the scattering. A model for the two-vertex interaction is needed to extract physical information from measured DCE cross sections. At the present time, a complete and coherent theory of such an interaction does not exist to the best of our knowledge. In a simple model, one can assume that the DCE process is just a second order charge exchange, where projectile and target exchange two uncorrelated isovector virtual mesons. The transition from initial $|i\rangle$ to final $|f\rangle$ reaction channels proceeds via the intermediate channels $|n\rangle$. Here the term channel is used to refer to a particular internal state of a partition in a particular state of relative motion [47]. This gives rise to a *VGV*-like term in the volume integral $\left|J_\alpha^{DCE}\right|^2$ which describes the action of the interaction *V* in two vertices. As pointed out in ref. [19], for DCE it has a non-vanishing contribution in a region around the overlapping surfaces of the colliding nuclei. The propagator is

$$G = \sum_n \frac{|n\rangle\langle n|}{E_n - (E_i + E_f)/2} \qquad (7)$$

where $E_{i,n,f}$ indicate the energies of the initial, intermediate and final channels, respectively. The explicit coordinate representation of *G*, which accounts for the relative motion in $|n\rangle$ is given in ref. [47]. In eq. (7), $E_n$ is a complex number whose imaginary component represents the off-shell propagation through the virtual intermediate states. This approach is analogous to the double-phonon model in giant resonance studies [48].

## 4. The experiment

The $^{40}$Ca($^{18}$O,$^{18}$Ne)$^{40}$Ar DCE reaction has been measured at the INFN-LNS laboratory in Catania together with the competing processes: $^{40}$Ca($^{18}$O,$^{18}$F)$^{40}$K single charge exchange, $^{40}$Ca($^{18}$O,$^{20}$Ne)$^{38}$Ar two-proton (2p) transfer and $^{40}$Ca($^{18}$O,$^{16}$O)$^{42}$Ca two-neutron (2n) transfer. A beam of $^{18}$O$^{4+}$ ions, extracted by the K800 Superconducting Cyclotron accelerator, bombarded a 279±30 μg/cm$^2$ Ca target, at 270 MeV incident energy. A total charge of 3.6 mC was integrated by a Faraday cup, downstream the target. The ejectiles produced in the collisions were momentum-analysed by the MAGNEX large acceptance spectrometer [49] and detected by its focal plane detector [50], [51]. An angular range of -1.2° < $\theta_{lab}$ < +8° in the laboratory frame was explored, corresponding to scattering angles in the center of mass 0° < $\theta_{CM}$ < 12°. The ejectiles identification was achieved as described in refs. [52], [53]. The positions and angles of the selected ions measured at the focal plane were used as input for a 10$^{th}$ order ray-reconstruction of the scattering angle $\theta_{CM}$ and excitation energy $E_x = Q_0 - Q$ (where $Q_0$ is the ground-to-ground state reaction *Q*-value) [54], [55], [56]. Figure 1 shows examples of the measured energy spectra. An energy resolution of ~500 keV (full width at half maximum) is obtained similarly to ref. [57]. The absolute cross section was extracted from measured yields according to ref. [54]. A systematic error of ±10% was estimated from the uncertainty in the target thickness and beam collection.

## 5. Experimental results

In the $^{40}$Ca($^{18}$O,$^{20}$Ne)$^{38}$Ar 2p-transfer energy spectrum of Figure 1a, the cross section tends to increase with excitation energy as a consequence of the kinematical *Q*-matching conditions ($Q_{opt}$ = 32 MeV). Known low-lying states are identified indicating the suppression of low multipolarity transitions due to *L*-matching conditions ($L_{opt}$ = 6). The *L*- and *Q*-optimum for the second step 2n-transfer $^{38}$Ar($^{20}$Ne,$^{18}$Ne)$^{40}$Ar are similar. Thus multistep transfer reactions are expected to be strongly suppressed in the population of the mismatched (*L* = 0, *Q* = -2.9 MeV) $^{40}$Ar ground state. In addition, the required condition of a 2n-transfer from the high spin intermediate states populated in the $^{40}$Ca($^{18}$O,$^{20}$Ne)$^{38}$Ar to the 0$^+$ $^{40}$Ar$_{g.s.}$ gives a supplementary reduction due to the vanishing Clebsch-Gordan coefficients. The cross section around zero-degree is ~3 µb/sr for the $^{40}$Ca($^{18}$O,$^{20}$Ne)$^{38}$Ar$_{gs}$, not larger than the $^{40}$Ca($^{18}$O,$^{18}$Ne)$^{40}$Ar$_{gs}$ (~ 11 µb/sr). This is very different from what reported in $^{14}$C + $^{40}$Ca at 51 MeV where the $^{40}$Ca($^{14}$C,$^{16}$O)$^{38}$Ar 2p-transfer cross section is ~1 mb/sr, i.e. almost two orders of magnitude larger than the corresponding $^{40}$Ca($^{14}$C,$^{14}$O)$^{40}$Ar DCE [18]. Bes et al. [19] and Dasso and Vitturi [20] conclude that the $^{14}$C + $^{40}$Ca → $^{16}$O + $^{38}$Ar → $^{14}$O$_{gs}$ + $^{40}$Ar$_{gs}$ transfer route is the leading mechanism feeding the $^{40}$Ar$_{gs}$. The reason is the much better matching of the 2p-transfer in the $^{14}$C + $^{40}$Ca ($Q_{opt}$ = 10 MeV, $L_{opt}$ = 3) compared to our case. Assuming a similar scaling between 2p-transfer and DCE for the present data, an upper limit of 30 nb/sr in the ($^{18}$O,$^{18}$Ne) reaction channel is estimated for the $^{18}$O + $^{40}$Ca → $^{20}$Ne + $^{38}$Ar → $^{18}$Ne$_{gs}$ + $^{40}$Ar$_{gs}$ multi-step route. Even considering the coherent sum of DCE and transfer amplitudes, a contribution of less than 10% is found at zero-degree.

The 2n-pickup 2p-stripping channel $^{18}$O + $^{40}$Ca → $^{16}$O + $^{42}$Ca → $^{18}$Ne$_{gs}$ + $^{40}$Ar$_{gs}$ is unlikely to contribute significantly, since the first step is already very suppressed in our experiments, with cross sections which are about half the cross section of the 2p-transfer.

The $^{40}$Ca($^{18}$O,$^{18}$F)$^{40}$K single charge exchange spectrum is shown in Figure 1b. Some structures are observed below 5 MeV excitation energy, however the limited resolution and the high level density do not allow to isolate single transitions. The strongest group is between 0.5 MeV and 1.2 MeV where the transitions to the known 2$^-$ and 5$^-$ states of $^{40}$K at 0.800 and 0.892 MeV and those to the excited states of the $^{18}$F ejectiles at 0.937, 1.041, 1.080 and 1.121 MeV, if populated, are not resolved [58], [59]. In particular, the dominance of the excited states of $^{18}$F at 1.041, 1.080 and 1.121 MeV is ruled out by a least square analysis, considering that they will generate Doppler broadened peaks with an extra width of about 300 keV. A number of $^{40}$K states are known in the region between 1.8 and 2.8 MeV. Calculations based on the Quasi Particle Random Phase Approximation – Distorted Wave Born Approach of ref. [45] indicate that the cross-section is mainly distributed among the 4$^-$, 2$^-$, 1$^+$ and 3$^-$ transitions. In particular the 1$^+$ accounts for about 40 µb/sr, consistent with the 38 µb/sr extracted from eqs. (3)-(4) using the parameters values reported in Section 6.1. A more detailed analysis of the single charge exchange reaction is beyond the scope of the present paper and will be published elsewhere.

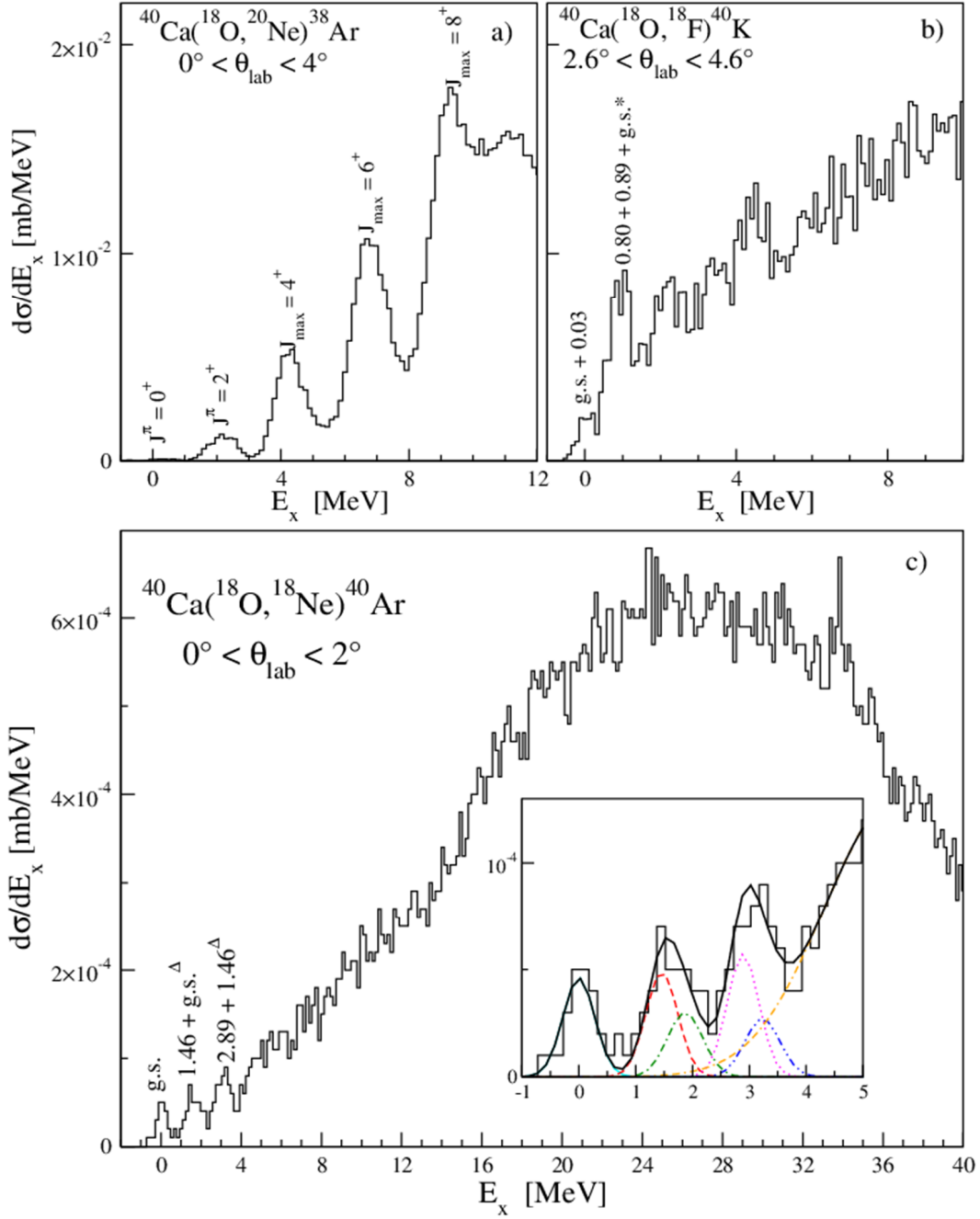

Figure 1. (a) Energy spectrum measured in the $^{40}$Ca($^{18}$O,$^{20}$Ne)$^{38}$Ar 2p-transfer. Above 3 MeV excitation energy, different states are overlapped in the observed peaks and the maximum angular momentum ($J_{max}$) is indicated according to [60]. (b) Energy spectrum from $^{40}$Ca($^{18}$O,$^{18}$F)$^{40}$K single charge exchange. The symbol g.s.* indicates the $^{40}$Ca($^{18}$O,$^{18}$F_{0.937MeV})^{40}$K$_{g.s.}$ transition. (c) Energy spectrum from $^{40}$Ca($^{18}$O,$^{18}$Ne)$^{40}$Ar DCE. The symbols g.s.$^{\Delta}$ and 1.46$^{\Delta}$ indicate the $^{40}$Ca($^{18}$O,$^{18}$Ne_{1.87MeV})^{40}$Ar$_{g.s.}$ and $^{40}$Ca($^{18}$O,$^{18}$Ne_{1.87MeV})^{40}$Ar$_{1.46MeV}$ transitions, respectively. In the insert, a zoomed view of the low-lying states and, superimposed (black solid line), a fit with 6 Gaussian functions are shown. They are centered at 0 (cyan solid), 1.46 (red dashed), 1.87 (green dot-dashed), 2.89 (magenta dotted), (1.46 + 1.87) = 3.33 (blue double dot-dashed), 5.6 MeV (orange dot-double dashed). The widths are given by the experimental resolution plus the Doppler broadening, except for the 5.6 MeV Gaussian whose width is 3 MeV. In (b) and (c) the symbol + indicates the presence of unresolved states.

In the DCE energy spectrum of Figure 1c, the $^{40}$Ar ground state is clearly separated from the not resolved doublet of states $^{40}$Ar $2^+$ at 1.460 MeV and $^{18}$Ne $2^+$ at 1.887 MeV. At higher excitation energy the measured yield is spread over many overlapping states. The angular distribution for the transition to the $^{40}$Ar $0^+$ ground state is shown in Figure 2. A clear oscillating pattern is observed. The position of the minima is well described by a $|j_0(qR)|^2$ Bessel function, where $R = 1.4$ $(A_1^{1/3}+A_2^{1/3})$ and $A_{1,2}$ is the mass number of projectile and target. Such an oscillating pattern is not expected in complex multistep transfer reactions, due to the large number of angular momenta involved in the intermediate channels, which would determine a structure-less cross section slowly decreasing at larger angles. The experimental slope is shallower than the Bessel function as expected since a plane-wave description is not appropriate [47]. Despite that, a very simple model of $L = 0$ direct process reasonably well describes the main features of the experimental angular distribution.

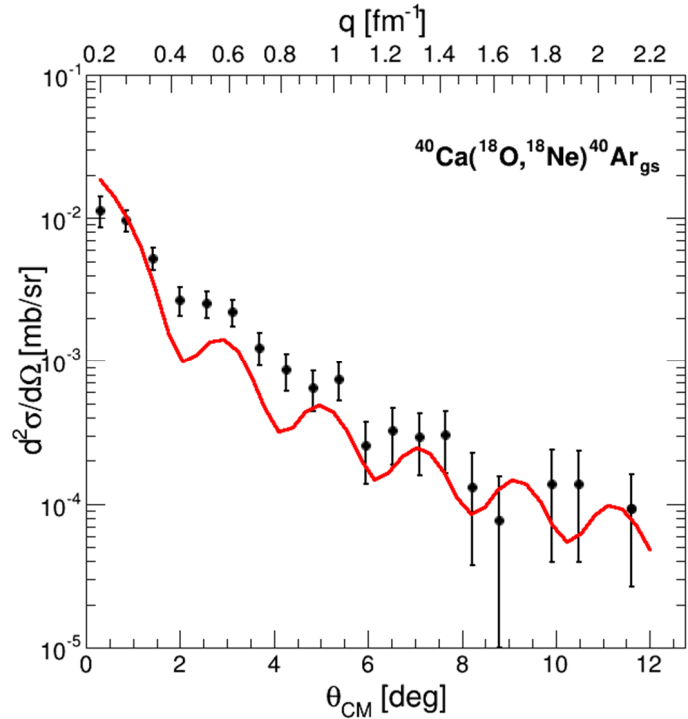

Figure 2. Differential cross section of the $^{40}$Ca($^{18}$O,$^{18}$Ne)$^{40}$Ar$_{g.s.}$ transition as a function of $\theta_{CM}$ and $q$. The error bars include a statistical contribution and a component due to the solid angle determination. The red curve represents the $L = 0$ Bessel function folded with the experimental angular resolution (~ 0.6°) and scaled to reproduce the incoherent sum of the predicted double $F$ and $GT$ cross sections (see Section 6.3).

## 6. Cross section analysis

DCE cross section factorization, described by eqs. (5-7), distinguishes between transitions originated by spin-isospin ($GT$-like) and isospin ($F$-like) operators trough the volume integral of nucleon-nucleon isovector interaction. Different multipolarities are allowed, each one characterized by an energy distribution which enters in the expression of the propagator. The sum over the many involved multipoles is important in DCE reactions, due to the large amount of momentum available, similarly to the $0\nu\beta\beta$ case (see for example Ref. [61]). In the case that enough excitation energy is available to populate all the channels of the intermediate partition relevant for the DCE process, a closure approximation can be considered, similarly to what is often done for *0νββ* analyses. In this case the the propagator of eq. (7) simplifies, since $\sum_n |n\rangle\langle n| = 1$ and the energy denominator is replaced by an average energy parameter. Assuming this approximation we found that the *F*-like and *GT*-like NMEs change of less than 20% for average energy ranging between 0 and 50 MeV. This indicates that the NMEs extracted using eqs. (5-7) are not much affected by changes of the energy distribution in the intermediate partition. The reason is that the kinetic energy in the propagator is much larger than the excitation energy. The overall effect of the higher multipolarities in the propagator is expected of the same order, since they globally determine a change of the average of the energy distribution within the explored range. This does not mean that higher multipole accounts for a small part of the strength, but that the overall effect in the NME of the redistribution of the strength to higher energies and multipoles is not necessarily large.

In order to check the validity of the factorization in eqs. (5-7) with existing data on β-decay strength distributions we deduce in section 6.1 and 6.2 the unit cross section and the matrix element for the DCE process assuming either a pure double *GT* or *F* transition, respectively.

*6.1 Gamow-Teller*

As discussed in Section 3, the $J_{GT}^{DCE}$ volume integral for the $^{18}$O + $^{40}$Ca → $^{18}$F$_{gs}$(1$^+$) + $^{40}$K$_{0-8\text{MeV}}$(1$^+$) → $^{18}$Ne$_{gs}$ + $^{40}$Ar$_{gs}$ at 270 MeV incident energy is estimated starting from the CE volume integral. We get $J_{GT}$ = 231 MeV·fm$^3$ using the isovector parts of the D3Y G-matrix [62], which includes spin-dependent and spin-independent direct and exchange central interactions. This is known to be an adequate choice at the energy of the present experiment, as described in [45]. The *G* propagator of eq. (7) is calculated summing over the on-shell energy distribution of $^{40}$K 1$^+$ states observed in high resolution ($^3$He,t) data on $^{40}$Ca target [63] and on $^{18}$F 1$^+$ ground state, as sketched in Figure 3. The off-shell contribution to *G*, accounting for ~7 MeV (full width at half maximum), is estimated from the crossing time (Δt ~ 27 fm/c) of the two pairs of nucleons participating in the DCE at 15 AMeV, assuming a correlation length of 4.8 fm from ref. [64].

A distortion factor $N_\alpha^D$ ~ 0.042 is calculated as the ratio of Distorted Wave over Plane Wave CE cross sections using the double folded optical potential of ref. [45]. Taking into account the kinematic factor *K* ~ 0.0089 [37], a $\hat{\sigma}_{GT}^{DCE}$ ~ 76 μb/sr is estimated for the $^{40}$Ca($^{18}$O,$^{18}$Ne)$^{40}$Ar DCE reaction at *q* = 0 from eq. (6). The correction factor of eq. (5) for *q* = 0.18 fm$^{-1}$, corresponding to $\theta_{CM}$ ~ 0° (measured angular interval 0° < $\theta_{CM}$ < 0.6°) is $F_{GT}^{DCE}$ ~ 0.72. From the measured $\frac{d\sigma^{DCE}}{d\Omega}$ ($\theta_{CM} = 0°$) = 11 μb/sr, one obtains an estimation of the maximum strength from eq. (5)

$$B^{DCE}(2GT) = B_T^{DCE}(2GT) \cdot B_P^{DCE}(2GT) = \frac{\frac{d\sigma^{DCE}}{d\Omega}(q,E_x)}{\hat{\sigma}_{GT}^{DCE}(E_p,A) F_{GT}^{DCE}(q,E_x)} \leq 0.20. \qquad (8)$$

This is compared to the value obtained combining the strengths, taken from literature, for the transitions in the projectile and target sketched in Figure 3 and listed in Table 1

$$B(2GT) = B_P(2GT) \cdot B_T(2GT) = 0.11 \qquad (9)$$

Here the $B$ transition strengths reduced for spin and isospin according to ref. [40] are used. In particular, for the projectile we have

$$B_P(2GT) = B[GT;{}^{18}O_{gs}(0^+) \rightarrow {}^{18}F_{gs}(1^+)] \cdot B[GT;{}^{18}F_{gs}(1^+) \rightarrow {}^{18}Ne_{gs}(0^+)] = 1.14 \qquad (10)$$

where only the population of the $^{18}F_{g.s.}$ is taken into account, as found in ref. [65] and the $B(GT)$ for the second step is from [66]. These factors are listed in table 1.

For the target

$$B_T(2GT) = \sum (B[GT;{}^{40}Ca_{gs}(0^+) \rightarrow {}^{40}K(1^+)] \cdot B[GT;{}^{40}K(1^+) \rightarrow {}^{40}Ar_{gs}(0^+)]) = 0.095 \qquad (11)$$

where the sum refers to the transitions to the $^{40}$K $1^+$ states up to 8 MeV observed in ref. [67]. The $B[GT;{}^{40}K(1^+) \rightarrow {}^{40}Ar_{gs}(0^+)]$ are taken from ref. [67]. For the $^{40}Ca_{gs}(0^+) \rightarrow {}^{40}K(1^+)$ transitions, we use high resolution $^{40}$Ca($^3$He,t)$^{40}$Sc data not yet published from Fujita et al. [63], assuming isospin symmetry. These data were also compared to previous results taken from literature: the $^{40}$Ca(p,n)$^{40}$Sc reaction at 159 MeV from [36] and at 134 MeV from [68]. In ref. [36] Taddeucci et al. give zero-degree cross section 1.2 mb/sr and a value of $B(GT; 2.73\ \text{MeV}) = 0.21 \pm 0.04$ for the transition to the strongest $1^+$ state of $^{40}$K at 2.73 MeV. These results conflict with Chittrakarn et al. [68] who measure a zero-degree cross section of 0.48 mb/sr for the same state, from which one can extract $B(GT; 2.73\ \text{MeV}) \sim 0.084$. In addition we also considered the results of Park et al. [69] who extract $B(GT; 2.73\ \text{MeV}) = 0.14 \pm 0.02$ from multiple decomposition analysis of the zero-degree cross section of the $^{40}$Ca(n,p)$^{40}$K reaction at 170 MeV. However, the results of Park et al. could be influenced by large systematic errors, due to the poor energy resolution and the uncertainties of the multiple decomposition analysis.

Preliminary results from the high resolution experiment of Fujita et al. [63] confirm the results of Chittrakarn et al. and this gives us confidence about the results from zero-degree cross section of the ($^3$He,t) reaction. In addition the ($^3$He,t) experiment also shows that, apart the $1^+$ state at 2.73 MeV, which carries a strength of $B(GT; 2.73\ \text{MeV}) = 0.069 \pm 0.006$, the $GT$ strength is fragmented in other 10 satellites. The two largest are at 2.33 and 4.40 MeV, which account for $B(GT; 2.33\ \text{MeV}) = 0.014 \pm 0.001$ and $B(GT; 4.40\ \text{MeV}) = 0.018 \pm 0.002$, respectively.

Only the 2.33, 2.73 and 4.40 MeV $1^+$ states of $^{40}$K give a not negligible contribution to the eq. (11). Merging the $B(GT)$ values from [67] and [63], we get the sum given in eq. (11). The total results and the partial numbers used to get this estimation are listed in Table 1.

The small value of $B^{DCE}(2GT)$ and $B(2GT)$ for the $^{40}$Ca is a consequence of the Pauli blocking in this doubly magic system. The $GT$ transitions take place only through the small $1f_{7/2}$, $1f_{5/2}$ particle and $1d_{3/2}$ hole components of the $^{40}$Ca$_{g.s.}$ wave function, which account for about 14% of the total [70], [71].

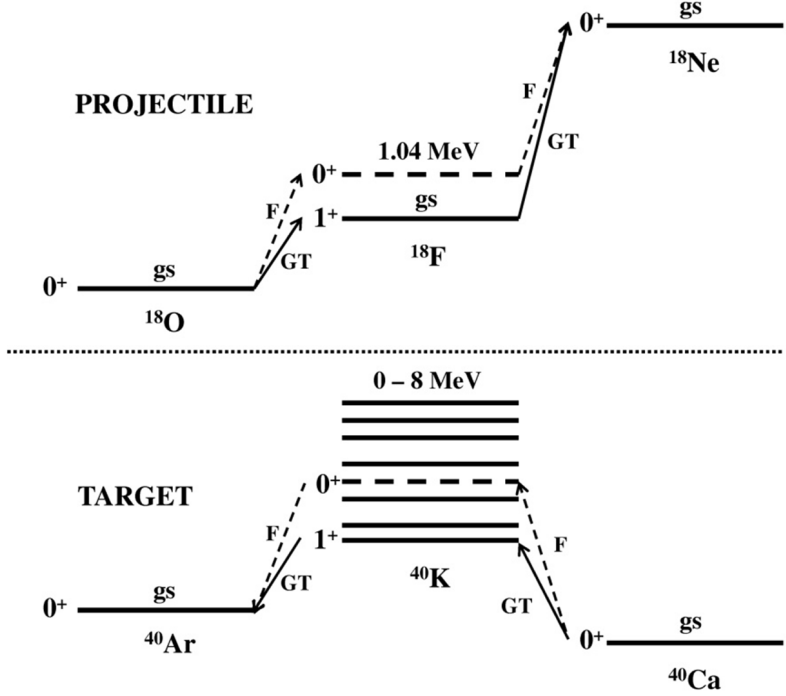

Figure 3. Diagram for the *GT* and *F* projectile and target transitions used for the determination of the *B*(2*GT*) and *B*(2*F*). See text.

*6.2  Fermi*

A similar procedure was applied assuming a pure double Fermi operator for the DCE to the $^{40}$Ar$_{gs}$. At the present energy the CE volume integral is $J_F = 253$ MeV·fm$^3$, very close to the *GT* case. Only the $^{40}$K 0$^+$ state at 4.38 MeV and the $^{18}$F 0$^+$ state at 1.04 MeV are considered in the intermediate channel. By the same arguments of the *GT* case, we obtain $\hat{\sigma}_F^{DCE} \sim 46$ μb/sr from eq. (6) and $F_F^{DCE} \sim 0.77$ at $\theta_{CM} \sim 0°$. As a consequence

$$B^{DCE}(2F) = B_T^{DCE}(2F) \cdot B_P^{DCE}(2F) = \frac{\frac{d\sigma^{DCE}}{d\Omega}(q,E_x)}{\hat{\sigma}_F^{DCE}(E_p,A) F_F^{DCE}(q,E_x)} \leq 0.31. \tag{12}$$

This value is close to the product of the known *B*(*F*) for the transitions in the projectile and target through the 1.04 MeV and 4.38 MeV 0$^+$ states of $^{18}$F and $^{40}$K, respectively (see Fig. 3):

$$B(2F) = B_P(2F) \cdot B_T(2F) = 0.42 \tag{13}$$

Here $B_P(2F) = 4$ is taken from the Fermi sum rule. $B_T(2F) = 0.053 \cdot 2$ is extracted by [67] and [63].

**Table 1**: Extracted strengths for pure Fermi and pure Gamow-Teller transitions and comparison with literature.

| α | $B^{DCE}(2\alpha)$ this work | $B(2\alpha)$ literature | $B_P$ ($^{18}O \rightarrow {}^{18}F$) | $B_P$ ($^{18}F \rightarrow {}^{18}Ne$) | $B_P(2\alpha)$ | $E(^{40}K)$ MeV | $B_T$ ($^{40}Ca \rightarrow {}^{40}K$) | $B_T$ ($^{40}K \rightarrow {}^{40}Ar$) | $B_T(2\alpha)$ |
|---|---|---|---|---|---|---|---|---|---|
| GT | 0.20 | 0.11 | 1.07[a] | 1.05[b] | 1.14 | 2.33<br>2.73<br>4.40<br>4.50-8 | 0.014[c]<br>0.069[c]<br>0.018[c]<br>8*0.007[c] | 1.03[d]<br>0.95[d]<br>0.54[d]<br>0.10[d] | 0.095 |
| F | 0.31 | 0.42 | 2 | 2 | 4 | | 0.053[c] | 2 | 0.106 |

a) From ref. [65], divided by $B_{(p,n)}(GT) = 3.049$
b) From ref. [66], divided by $B_{(p,n)}(GT) = 3.049$
c) From ref. [63]
d) From ref. [67]

### *6.3 DCE nuclear matrix elements*

Under the hypothesis of pure *GT* transition, $B_T^{DCE}(2GT)$ can be extracted dividing eq. (8) by $B_P^{DCE}(2GT)$, assuming $B_P^{DCE}(2GT) = B_P(2GT) = 1.14$ (see eq. (10)). The NME can be then derived from $B_T^{DCE}(GT)$ via

$$\left|M_T^{DCE}(\alpha)\right|^2 = B_T^{DCE}(\alpha) \qquad (14)$$

obtaining $M_T^{DCE}(GT) = 0.42 \pm 0.21$. In the lack of theoretical predictions for the $^{40}$Ca NME, it is worth to compare this value with the well studied $^{48}$Ca NME, which ranges from 0.67 [1] to 1.53 [4] depending on different models. The smaller value obtained for $^{40}$Ca is compatible with the effect of Pauli blocking present only in this system, which determines reduction of a factor about 7 for *F* and *GT*.

Analogously, in the case of pure Fermi process, we extract $M_T^{DCE}(F) = 0.28 \pm 0.14$ from eqs. (12-14). The systematic error in the determination of $M_T^{DCE}$ is about ±50%, estimated by checking the sensitivity of the results to the used parameters. It mainly comes from the uncertainty on volume integrals of the effective two-body interaction. The contribution of the experimental error (±10% systematic, ±25% statistical) is less relevant in this case.

Both *F* and *GT* contribute to the total cross section at $\theta_{CM} = 0°$. Their size can be predicted by $B(2GT) \cdot \widehat{\sigma}_{GT}^{DCE} \cdot F_{GT}^{DCE} \sim 6$ μb/sr for Gamow-Teller and $B(2F) \cdot \widehat{\sigma}_F^{DCE} \cdot F_F^{DCE} \sim 15$ μb/sr for Fermi. The comparison is much more accurate than the single estimation due to the common assumptions done. The *L* = 0 Bessel function shown in Figure 2 is scaled to give a cross section of 21 μb/sr at $\theta_{CM} = 0°$, which is the incoherent sum of the predicted *GT* and *F* cross sections. The comparison with the experimental data show a remarkable quantitative agreement. However, the effects of the interference should be studied in detail. The fact that both pure *GT*- and *F*-like extreme models give comparable contributions to the final cross section is a direct consequence of the similar volume integrals for both operators. The relation between these volume integrals resembles that for nucleon-nucleon interaction at 15 MeV. This indicates that the reaction mechanism is largely determined by the effective nucleon-nucleon interaction. Experiments at different incident energies are envisaged in order to explore conditions characterized by different weights of *GT*-like and *F*-like contributions and disentangle the role of each operator. In particular, higher energies are more suitable to study the *GT*-like NMEs, which are expected to be dominant in most of the *0νββ* decays.

### *6.4    0νββ nuclear matrix elements*

In the previous sections we showed that in the extreme hypothesis of pure Gamow-Teller-like or Fermi-like transition the extracted matrix elements are $M_T^{DCE}(GT)= 0.42 \pm 0.21$ or $M_T^{DCE}(F) = 0.28 \pm 0.14$, respectively. We notice that they are very similar, so even the weighted average, representing a more realistic combination of both contribution, will be. Assuming the known *GT* and *F* strengths from literature (see discussion section 6.3) we can get an estimate of the weights through the expression of the DCE cross section expected at zero-degree:

$$\frac{d\sigma}{d\Omega}(\theta = 0°, E_x = 0) = \hat{\sigma}_{GT}^{DCE} F_{GT}^{DCE} B(2GT) + \hat{\sigma}_F^{DCE} F_F^{DCE} B(2F) = 6\frac{\mu b}{sr} + 15\frac{\mu b}{sr} = 21\frac{\mu b}{sr} \quad (15)$$

From eq. (15) we have $\sqrt{6/21}$ for the *GT* and $\sqrt{15/21}$ for the *F* weights, respectively. The matrix elements weighted in this way are $M'^{DCE}_T(GT) = 0.22$ and $M'^{DCE}_T(F) = 0.24$. Consequently, we could infer the matrix element for the *0νββ* decay of $^{40}$Ca

$$M(0\nu\beta\beta; {}^{40}\text{Ca}) = [(g_v/g_a)^2 M'^{DCE}(F) + M'^{DCE}(GT)] = 0.62 \cdot 0.24 + 0.22 = 0.37 \pm 0.18 \quad (16)$$

where $g_{a,v}$ are the axial and vector coupling constants of the weak interaction, respectively [4]. This small number reflects the Pauli blocking, as discussed in Section 6.1.

To speculate, a comparison between the present result for $^{40}$Ca and the NME of *0νββ* decay of $^{48}$Ca can be done assuming pure *F* and *GT* and artificially removing the effect of the Pauli blocking, since the same single particle shells are involved but no Pauli blocking is active in the $^{48}$Ca case. This is possible by just multiplying $M(0\nu\beta\beta; {}^{40}\text{Ca}) \cdot 7 = 2.6 \pm 1.3$. It is noteworthy that this number is compatible with literature for the calculations of the $^{48}$Ca *0νββ* NME [3], [10].

## 7. Conclusions

In conclusion, this work shows for the first time high resolution and statistically significant experimental data on heavy-ion double charge exchange reactions in a wide range of transferred momenta. The measured cross-section angular distribution shows a clear oscillating pattern, remarkably described by an *L* = 0 Bessel function, indicating that a simple mechanism is dominant in the DCE reaction. This is confirmed by the observed suppression of the multi-nucleon transfer routes.

Strengths factors and matrix elements are extracted under the hypothesis of a two-step charge exchange process. Despite the approximations used in our model, which determine an uncertainty of ±50%, the present results are compatible with the values known from literature, signaling that the main physics content has been kept. The DCE unit cross section is likely to be a predictable quantity, in analogy to the CE processes. We believe that this finding is mainly due to the particularly simple transitions which take place in the $^{18}O \rightarrow {}^{18}F \rightarrow {}^{18}Ne$, characterized by a strong dominance of single $1^+$ and $0^+$ $^{18}F$ states in both *GT* and *F* transitions, respectively. This makes the ($^{18}O$,$^{18}Ne$) reaction very interesting to investigate the DCE response of the nuclei involved in *0νββ* research.

A deeper investigation of DCE reactions is worthwhile in the future, studying other systems and different bombarding energies, in order to explore the systematic behavior. In all cases the contextual measurements of the multi-nucleon transfer and single charge exchange channels is mandatory. A rigorous treatment of the process in a multistep direct reaction quantum-mechanichal framework will be the next step toward a more accurate determination of *0νββ* NMEs.

**Acknowledgements**

The authors wish to thank Prof. Y. Fujita for his kindness to provide us useful information and Prof. H. Lenske for the fruitful discussions on reaction theory. The authors acknowledge also the staff of the LNS accelerator division for the excellent beam delivered.

**Bibliography**

[1]   E. Caurier, J.Menendez, F. Nowacki, A. Poves, Phys. Rev. Lett. 100 (2008) 052503.

[2]   J. Suhonen and M.Kortelainen, Int. Journ. of Mod. Phys. E 17 (2008) 1.

[3]   N. L. Vaquero, T.R. Rodriguez, J. L.Egido, Phys. Rev. Lett. 111 (2013) 142501.

[4]   J. Barea, J. Kotila, F. Iachello, Phys. Rev. C 87 (2013) 014315.

[5]   H. Akimune, et al., Phys. Lett. B 394 (1997) 23.

[6]   J. P. Schiffer, et al., Phys. Rev. lett. 100 (2008) 112501.

[7]   D. Frekers, Prog. Part. Nucl. Phys. 64 (2010) 281.

[8]   C.J. Guess, et al., Phys. Rev. C 83 (2011) 064318.

[9]   S. J. Freeman and J. P. Schiffer, J. Phys. G: Nucl. Part. Phys. 39 (2012) 124004.

[10] J. Barea, J. Kotila and F. Iachello, Phys. Rev. Lett. 109 (2012) 042501.

[11] Report to the Nuclear Science Advisory Committee, Neutrinoless Double Beta Decay, 2014.

[12] J.D. Vergados, Phys. Rev. D 25 (1982) 914.

[13] A. Fazely and L. C. Liu, Phys. Rev. Lett. 57 (1986) 968.

[14] S. Mordechai, et al., Phys. Rev. Lett. 61 (1988) 531.

[15] N. Auerbach, et al., Ann. Phys. 192 (1989) 77.

[16] J. Blomgren, et al., Phys. Lett. B 362 (1995) 34.

[17] F. Naulin, et al., Phys. Rev. C 25 (1982) 1074.

[18] D. M. Drake, et al., Phys. Rev. Lett. 45 (1980) 1765.

[19] D.R. Bes, O. Dragun, E.E. Maqueda, Nucl. Phys. A 405 (1983) 313.

[20] C.H. Dasso, A. Vitturi, Phys. Rev. C 34 (1986) 743.

[21] W. von Oertzen, et al., Nucl. Phys. A 588 (1995) 129c.

[22] H. Matsubara, et al., Few-Body Systems 54 (2013) 1433.

[23] D. M. Brink, Phys. Lett. B 40 (1972) 37.

[24] P. Puppe, et al., Phys. Rev. C 84 (2011) 051305.

[25] H. Ejiri, Czech. J. Phys. 56 (2006) 459.

[26] J. Suhonen and M. Kortelainen Czech J. Phys. 56 (2006) 519.

[27] W.P. Alford and B.M. Spicer, Advances in Nucl. Phys. 24 (1998) 1.

[28] J. Menéndez, D. Gazit, A. Schwenk, Phys. Rev. Lett. 107 (2011) 062501.

[29] L. Mandelstam and Ig. Tamm, J. Phys. USSR 9 (1945) 249.

[30] Workshop on multistep direct reactions, edited by R.H. Lemmer, World Scientific, ISBN 981-02-1171-6 (1992).

[31] H. Feshbach, A. Kerman, S. Koonin, Ann. Phys. 125 (1980) 477.

[32] T. Tamura, T. Udagawa, H. Lenske, Phys. Rev. C 26 (1982) 379.

[33] H. Nishioka, H. A. Weidenmuller, S. Yoshida, Ann. Phys. 183 (1988) 166.

[34] A.J. Koning and J.M. Akkermans, Computer Physics Communications 85 (1995) 110.

[35] A. J. Koning and M. B. Chadwick, Phys. Rev. C 56 (1997) 970.

[36] T. N. Taddeucci et al. Phys. Rev. C 28 (1983) 2511.

[37] T.N. Taddeucci et al. Nucl. Phys. A469 (1987) 125.

[38] F. Osterfeld, Rev. of Mod. Phys. 64 (1992) 491.

[39] D. Frekers, et al., Nucl. Phys. A 916 (2013) 219.

[40] Y. Fujita, B. Rubio, W. Gelletly, Progress in Particle and Nuclear Physics 66 (2011) 549.

[41] C. Baumer, et al., Phys. Rev. C 71 (2005) 024603.

[42] A. Negret, et al., J. Phys. G: Nucl. Part. Phys. 31 (2005) 1931.

[43] C. Brendel, et al., Nucl. Phys. A 477 (1988) 162.

[44] H. Lenske, Nucl. Phys. A 482 (1988) 343c.

[45] F. Cappuzzello et al., Nuclear Physics A 739 (2004) 30.

[46] A. Etchegoyen, et al., Phys. Rev. C 38 (1988) 2124.

[47] G. R. Satchler, Direct Nuclear reactions, Oxford Science Publications, 1983.

[48] Ph. Chomaz, N. Frascaria, Phys. Reports 252 (1995) 275.

[49] F. Cappuzzello et al., MAGNEX: an innovative large acceptance spectrometer for nuclear reaction studies in: Magnets: Types, Uses and Safety, Nova Publisher Inc., New York, 2011, pp 1-63.

[50] M. Cavallaro, et al., Eur. Phys. J. A (2012) 48: 59.

[51] D. Carbone, et al., Eur. Phys. J. A (2012) 48: 60.

[52] M. Bondì, et al., AIP Conf. Proc. 1595 (2014) 245.

[53] F. Cappuzzello, et al., Nucl. Instr. and Meth. A 621 (2010) 419.

[54] M. Cavallaro, et al., Nucl. Inst. and Meth. A 648 (2011) 46.

[55] F. Cappuzzello, et al., Nucl. Instr. and Meth. A 638 (2011) 74.

[56] M. Cavallaro, et al., Nucl. Instr. and Meth. A 637 (2011) 77.

[57] F. Cappuzzello, et al., Nucl. Instr. and Meth. A 763 (2014) 314.

[58] F. Ajzenberg-Selove, et al., Phys. Rev. C 32 (1985) 756.

[59] D. R. Tilley, et al., Nucl. Phys. A 595 (1995) 1.

[60] J. A. Cameron and B. Singh, Nucl. Data Sheets 109 (2008) 1.

[61] J. Hyvarinen and J. Suhonen, Phys. Rev. C 91 (2015) 024613.

[62] F. Hofmann and H. Lenske, Phys. Rev. C 57 (1998) 2281.

[63] Y. Fujita private communication.

[64] M. Matsuo, Phys. Rev. C 73 (2006) 044309.

[65] D.J.Mercer et al. Phys. Rev. C 49 (1994) 3104.

[66] D. E. Alburger and D. H. Wilkinson, Phys. Lett. B 32 (1970) 190.

[67] M. Bhattacharya, C.D. Goodman and A. Garcia, Phys. Rev. C 80 (2009) 055501.

[68] T. Chittrakarn et al., Phys. Rev. C 34 (1986) 80.

[69] B. K. Park, et al., Phys. Rev. C 45 (1992) 1791.

[70] W. Tornow, et al., Phys. Rev. C 42 (1990) 693.

[71] S. Adachi, et al., Nucl. Phys. A 438 (1985) 1.